\documentclass[acmsmall,nonacm]{acmart}
\AtBeginDocument{%
  \providecommand\BibTeX{{%
    \normalfont B\kern-0.5em{\scshape i\kern-0.25em b}\kern-0.8em\TeX}}}

\setcopyright{acmcopyright}
\copyrightyear{2023}
\acmYear{2023}
\usepackage{algorithmic}
\usepackage{algorithm}
\usepackage{graphicx}
\usepackage[toc,page]{appendix}

\begin{document}

\title{Adaptive Compression-Aware Split Learning and Inference for Enhanced Network Efficiency}

\author{Akrit Mudvari}
\email{akrit.mudvari@yale.edu}
\orcid{XXXX-XXXX-XXXX}
\affiliation{%
  \institution{Yale University}
  \city{New Haven}
  \state{CT}
  \country{USA}
  \postcode{06511}
}

\author{Antero Vainio}
\affiliation{%
  \institution{University of Helsinki}
  \city{Helsinki}
  \country{Finland}
}
\email{antero.vainio@helsinki.fi}

\author{Iason Ofeidis}
\affiliation{%
  \institution{Yale University}
  \city{New Haven}
  \state{CT}
  \country{USA}
  \postcode{06511}
}
\email{iason.ofeidis@yale.edu}

\author{Sasu Tarkoma}
\affiliation{%
  \institution{University of Helsinki}
  \city{Helsinki}
  \country{Finland}
}
\email{sasu.tarkoma@helsinki.fi}

\author{Leandros Tassiulas}
\affiliation{%
  \institution{Yale University}
  \city{New Haven}
  \state{CT}
  \country{USA}
  \postcode{06511}
}
\email{leandros.tassiulas@yale.edu}

\renewcommand{\shortauthors}{Mudvari, et al.}

\begin{abstract}
The growing number of AI-driven applications in mobile devices has led to solutions that integrate deep learning models with the available edge-cloud resources. Due to multiple benefits such as reduction in on-device energy consumption, improved latency, improved network usage, and certain privacy improvements, split learning, where deep learning models are split away from the mobile device and computed in a distributed manner, has become an extensively explored topic.
Incorporating compression-aware methods (where learning adapts to compression level of the communicated data) has made split learning even more advantageous. This method could even offer a viable alternative to traditional methods, such as federated learning techniques. In this work, we develop an adaptive compression-aware split learning method ('deprune') to improve and train deep learning models so that they are much more network-efficient, which would make them ideal to deploy in weaker devices with the help of edge-cloud resources. This method is also extended ('prune') to very quickly train deep learning models through a transfer learning approach, which trades off little accuracy for much more network-efficient inference abilities. We show that the 'deprune' method can reduce network usage by 4x when compared with a split-learning approach (that does not use our method) without loss of accuracy, while also improving accuracy over compression-aware split-learning by 4 percent. Lastly, we show that the 'prune' method can reduce the training time for certain models by up to 6x without affecting the accuracy when compared against a compression-aware split-learning approach. 
\end{abstract}

\ccsdesc[500]{Computing methodologies~Machine learning}
\ccsdesc[300]{Networks~Network management}

\keywords{Deep Learning, Neural Network Splitting, Compression, Pruning, Transfer Learning, Augmented Reality, Internet of Things, Edge Computing, Fog Computing}


\maketitle

\section{Introduction}
The past decade or so has seen a meteoric rise in the scope and abundance of AI applications with deep learning as a core tool, leading to solutions or better performance in many areas such as speech recognition, computer vision and much more \cite{krizhevsky2017imagenet, he2016deep, girshick2015region, hinton2012deep, mikolov2013distributed}. It is estimated that the global AI market size was 120 billion USD in 2022 and is expected to reach 1.5 trillion dollars by 2030 \cite{PrecedenceResearch}, and even in each of our daily lives the impact of AI is becoming more and more pronounced, i.e., with the recent release of ChatGPT \cite{radford2018improving}. While beneficial, these methods, especially deep learning, tend to be quite expensive to execute. Supervised learning methods rely on huge datasets to learn, leading to a colossal demand for data acquisition and storage. The increasing demand of computational resources has also pushed the development of dedicated hardware platforms, i.e, Graphics Processing Units (GPUs) and Tensor Processing Units (TPUs), whose goal is to speed-up the training tasks. Another significant source of cost tends to be energy consumption in the devices that run the deep learning models, which not only poses an economic challenge but also raises other concerns such as energy scarcity and climate impacts \cite{nordgren2022artificial}. It has become quite necessary to envision and implement these deep learning models in settings that are optimized towards saving different forms of resources.

\begin{figure}[t]
  \centerline{
    \includegraphics[width=0.7\textwidth]{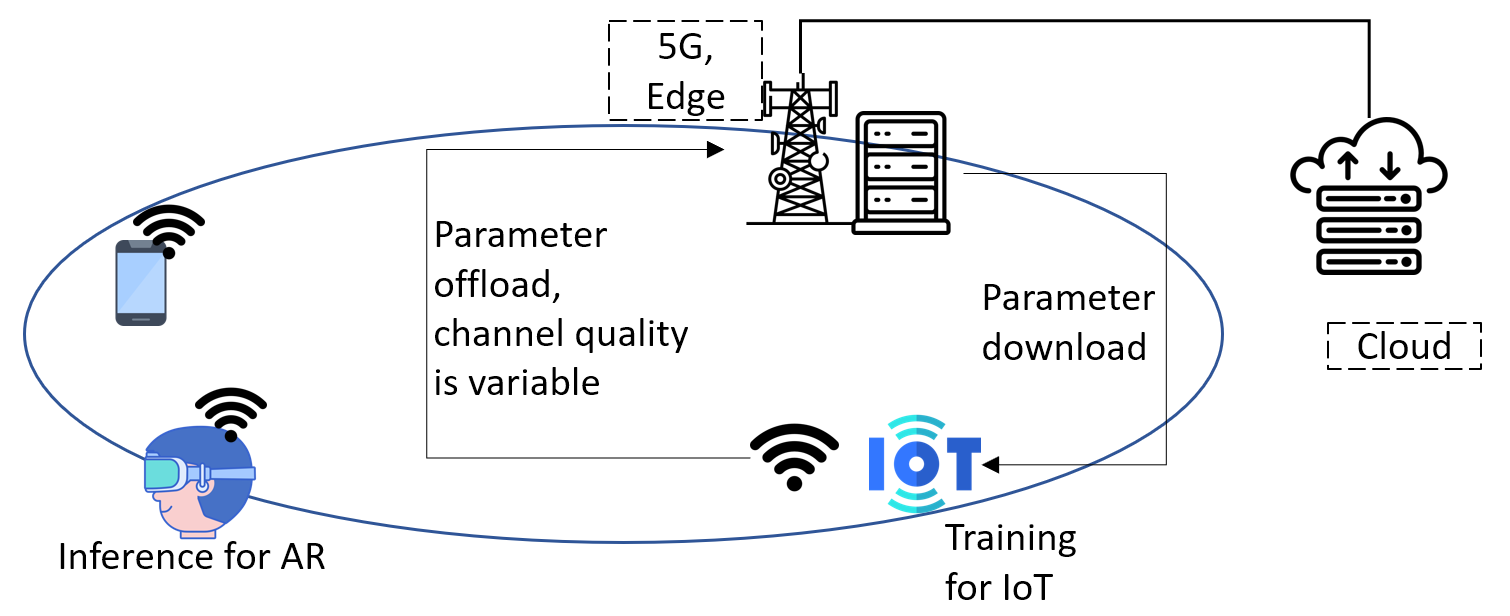}
  }
  \caption{Illustration of different weaker devices opting for nearby computation sources, for AI tasks} 
  \label{fig:illustration}
\end{figure}

There have been various efforts resulting in reduction of computational costs, i.e. through the implementation of smaller deep learning models \cite{fang2019tinier, womg2018tiny, jiao2019tinybert} and model pruning including numerous works on parameter pruning \cite{molchanov2019importance,carreira2018learning, zhao2019variational} as well as some works on feature pruning \cite{hu2016network, he2017channel, peng2019collaborative}, which allows for utilization of less GPUs and for shorter periods of time (for both inference as well as training). While this helps some devices reduce both energy costs as well as computational latency, it is not adequate for the vast majority of end devices like cell phones, laptops, IoT devices and AR/VR headsets to locally carry out the implementations \cite{huang2021proactive, liu2018edge}; as a result the better alternative for these devices tends to be utilizing another more powerful source of computation, i.e. via different forms of distributed learning. This will inevitably involve transferring certain amount of data over the communication networks to locations such as data centers, resulting in a higher network usage as well as introduction of a new source of latency. The ideal solution would therefore involve selecting a method that minimizes network usage and latency. 

Different strategies have been proposed for both training and inference of deep learning models in the networked environments. For instance, with the proliferation of data-driven services, such as targeted advertisement, personalized suggestion systems and more, it has become crucial to find an efficient method of conducting deep learning tasks that is both computation and communication efficient, but also has other features including privacy protection. While one could simply offload all the computation task to the nearby server for further processing, such an approach has a few problems and shortcomings, which split learning and inference are able to overcome. Split learning (at least as we define and discuss in this paper) refers to splitting a neural network into multiple partitions so that different parts of the deep learning model can be computed in different ways. For instance, consider a neural network with layers $l_1,….,l_m,….,l_n$ where $m<n$; here split learning can involve processing $l_1,….,l_m$ locally and then offloading $l_{m+1},….,l_n$ to a server for further computations. Such split learning methods can provide us with various benefits over simply offloading the model as a whole. First, unlike offloading raw data, we are able to send a processed output, which is able to remain privacy preserving \cite{jeong2018computation,wadhwa2023pfsl,yang2022differentially}. Second, the granularity provided by such division of task allows for more optimal scheduling and placement. Third, and perhaps more promising, as evidenced with recent research works \cite{wadhwa2023pfsl, arivazhagan2019federated}, it could also help with personalization of learned models by keeping the locally trained parts more personalized, and having a globally shared set of layers for aggregated learning. 

Certain other approaches allow for similar benefits in certain cases. Federated learning for instance has emerged as a popular and widely researched method of learning and aggregating the learned information in a privacy preserving manner. However, split learning has emerged as a promising alternative to federated learning \cite{singh2019detailed,thapa2022splitfed}. Split learning has been shown to perform better in numerous scenarios including when the number of model parameters is larger, or when more clients are involved \cite{singh2019detailed}. They are also more communication-efficient in many situations \cite{thapa2022splitfed} as compared with federated learning, and have the capability to be privacy preserving as raw data need not be offloaded. These reasons make exploration of split learning as a method of distributed deep learning quite important and promising. 

Previous works have explored split learning, i.e, in \cite{kang2017neurosurgeon, eshratifar2019bottlenet, shao2020bottlenet++}, which we will explore in the related works section in further details. Some of these works, for example in \cite{eshratifar2019bottlenet, shao2020bottlenet++}, have explored the feasibility of conducting split learning and inference in different networked scenarios, and have introduced state of the art compression methods to reduce network traffic to a great extend, while minimally affecting the model accuracy. However, while these advances are quite important for making split method useful for day-to-day applications, there are certain shortcomings, which need to be addressed before this method can become widely feasible. While network-efficient split methods can help conduct spit learning and inference, the most glaring weakness of current state-of-the-art approaches is that training has to happen again and again for different split configurations, specially for different compression levels. 
Various factors like network conditions, operation budget, latency constraints, and tolerance of service level agreement violation, result in the network operator having different preference when it comes to compression and accuracy trade off. Hence providing models suited to  different compression-accuracy trade offs/ compression levels becomes very time and resource consuming, or even outright impractical.    

In this work, we introduce a method that uses pruning and transfer learning to very quickly train deep learning models with different compression levels for the data being transferred during split learning, allowing for efficient creation of models suitable for different network conditions, computation resources, and service level agreements. This is done by introducing an encoder-decoder module with a method of training across different compression levels and at different positions of splitting. One use case of our approach would be providing AI-driven services like Augmented  Reality (AR) with latency constraints to handheld devices in a wireless environment. AI/AR applications requiring object detection may employ models like YOLO \cite{redmon2016you} or SSD \cite{liu2016ssd} but will not be able to meet the users' latency requirements \cite{FutureNetworks_2019}. With the models developed using our aforementioned approach, we could provide each user with different accuracy-compression trade offs, while split learning allows for optimal distribution of workload and an option to avoid offloading raw user data such as video feed or captured image. Hence, service level agreements can be met more efficiently. 

This encoder-decoder module, further explained in section \ref{sec:methods}, is a tool that uses the learned information from a model with certain compression level, and adaptively learns to optimize for a different compression level. As mentioned above, one of the two goals is to develop these approximately similar models suitable for different user requirements and environments; however, a modified version of the aforementioned approach is also very useful in a second way, where this encoder-decoder approach can be used to adaptively change compression configurations to train in a way where the training can result in effectively same level of accuracy at the end, while significantly cutting down on the network usage during the distributed training. By training on a high compression level, and then towards the end of the training process, adaptively switching to a lower compression level using the deprune method (discussed further in section \ref{sec:methods}), we observed that network usage could be reduced without significantly effecting the accuracy of the final learned model. Such training can be specially beneficial when weaker devices, i.e., at the edge of the network, have limited computation capability as well as high network cost; and since split learning is involved, raw data does not have to be offloaded.

\section{Related Work}
\label{sec:related}

Multitude of work have focused on extending placement and scheduling problems in an edge-cloud paradigm, allowing for a generic workload to be delegated to servers in a networked environment. For instance, in \cite{poularakis2020service}, any service that requires computational offloading is placed at the network edge and defined constraints include storage, computation and communication costs, and in \cite{poularakis2020approximation} the process is extended to data-intensive tasks. In \cite{pasteris2019service} and \cite{farhadi2021service}, algorithms are developed for service placement with guarantees for near optimal solutions, with \cite{pasteris2019service} creating a deterministic algorithm for general service placement and \cite{farhadi2021service} working on polynomial time solution for data-intensive tasks. As we can see in survey papers such as \cite{sonkoly2021survey, malazi2022dynamic, luo2021resource}, different types of placement and scheduling methods may be utilized for implementing state of the art techniques for efficient allocation of communication and computation resources in the edge-cloud paradigm. While these methods are an acceptable improvement over a naive deployment of deep learning models, for either training or inference, their generalized scope means that they fail to take into consideration unique features of deep learning models, or specific requirements of the clients, that may be leveraged to provide a more efficient allocation scheme for such implementations under an edge-cloud environment. And our developed methods can complementarily be implemented together with such placement and scheduling approaches. 

In \cite{murthy2016deep}, the authors develop a measurement-driven framework that decides which deep learning model to run and where, based on the different model-related metrics such as accuracy, frame rate of processed videos, energy consumption, and network utilization. In \cite{han2016mcdnn}, an approximate model scheduling is developed, where the trade-off between accuracy and resources is harnessed to optimize model accuracy. Various factors such as device energy, cloud costs and capacity, and execution deadline are taken as constraints, and an online variant of the method is also developed. Complementing the scheduling and placement research, there are methods of creating variants of the DNNs that aim to reduce space or computational demands, effectively creating a different NN such as through quantization of the entire network \cite{han2015deep}, binarization of the models \cite{rastegari2016xnor}, representation of the model as a lower rank representation after singular value decomposition \cite{xue2014singular}, whole model compression through algorithms consisting of low-rank tensor decomposition \cite{kim2015compression}, and more. 

Complementary to a lot of such efforts that treat the model as a whole and attempt to efficiently orchestrate NN model deployments, it is possible to add granularity to the methods and also potentially improve privacy \cite{jeong2018computation, yang2022differentially} by splitting the neural network into different parts for computation at different devices. While methods like deep leakage \cite{zhu2019deep} have haunted federated learning methods for a while now by eliminating privacy, no such method exists for reverse engineering split learning data to the best of our knowledge. Intuitively, there are a lot more parameters being updated in federated learning as opposed to the number of feature maps communicated during split learning, so the process developed to counter split learning is possibly a harder endeavor if not impossible.

In \cite{kang2017neurosurgeon}, the authors proposed a method for collaborating intelligence among the end devices and the mobile edge by dividing the neural network for partial computation at each end. Here, the authors investigated methods of discerning the best points of splitting and also showed that this method could be used to improve inference latency and energy efficiency of mobile devices for a wide range of deep learning models. This method is now well-understood as split learning (or vertical split learning, but we will use the term split learning to mean the vertical case). Capitalizing on this observation regarding the viability of splitting deep learning models as shown in \cite{kang2017neurosurgeon}, authors in \cite{eshratifar2019bottlenet} developed a compression aware training method for image classification neural networks that inserts a 'bottleneck structure' at the point of model splitting for efficient lossy compression. These 'bottleneck' structures use convolution methods to decrease the dimensions of the feature map for compression, and use de-convolution to obtain the larger image once the communication of compressed data has taken place. The authors show that lossy compression approach is complementary to lossless methods and also show great improvements in end-to-end latency and energy consumption. \cite{shao2020bottlenet++} improves on the compression structure proposed in \cite{eshratifar2019bottlenet}, specifically by considering different network channels. Here, they also demonstrate the superiority of such compression and decompression-based training method over the other methods including jpeg \cite{ko2018edge} method and quantization+Huffman \cite{shi2019improving} method, towards improving both communication overhead as well as transmission latency. \cite{matsubara2022bottlefit} proposes a method for specific classification tasks by running a generalized head network distillation in the first stage and then targeted knowledge distillation in the second state. Other works such as \cite{assine2021single, matsubara2020split, matsubara2021neural} have focused on object detection tasks, where the compression mechanism is analyzed for the object detection. While many of these efforts have focused on implementing a compression-aware split learning that reduces network usage by splitting at certain layer and compressing the data to be transferred, the glaring inadequacy in these methods is that every time a new compression level is needed, or if compression behavior needs to change, the training has to restart from the scratch. Once a compression factor has been decided on, i.e., by how much to compress, this value cannot dynamically change. This limits the scope of compression-aware training approaches since it would be unfeasible if not just inefficient to train from scratch for every compression configuration.   

In our work, we propose an adaptive compression-decompression module that creates approximate models for deep learning architectures suitable for split learning and inference, and we introduce two different novel algorithms, 'deprune' and 'prune', that greatly improve the performance during training as well as inference, i.e., by allowing us to train the deep learning model much more quickly and by improving models for faster inference. In the first step, we develop an adaptive compression-decompression method that allows for dynamic adjustment of compression levels through a 'filter' variable that we can change during training. In the second step, we introduce the 'deprune' method which uses this module to significantly reduce network usage during the training under a split-learning environment, while also reducing the latency. We also highlight the authenticity of this method by deploying it in networked testbeds. This approach is highly beneficial for IoT and other weaker end devices with AI tasks, as network usage and training latency can be significantly improved for split learning. Finally, we introduce the 'prune' method that can be used to transfer learned parameters to dynamically create deep learning models for inference that greatly outperform having to train from scratch, in that the training time is significantly reduced. Such a method can create approximate models for different network requirements that trade little accuracy for network performance, and help create deep learning models that the AI tasks can implement in a network constrained environment; i.e., a network manager may reduce performance for AI-driven XR tasks by selecting a model with higher compression level if network traffic becomes an issue.

\section{Proposed Methods}
\label{sec:methods}

In this section, we will describe the split learning architectures, including methods of training and implementing the algorithms. In subsection \ref{subsec:DePruning}, we propose a method where the network communication cost of training is reduced using our method. And in subsection \ref{subsec:Pruning}, we will describe a method of efficiently creating deep learning architectures that are able to quickly train and provide a set of models that offer a trade-off between accuracy and inference speed. But first, we will describe the formulation for deep learning architecture, as well as compression and decompression methods that we implement in our split-learning architecture, in subsection \ref{subsec:formulation}.

\subsection{Formulation}
\label{subsec:formulation}

Our implementation is designed for general deep learning models, and so we define a model $M$ with an arbitrary $L$ number of layers where the outputs of the layers are represented by $l = {l_0,l_1,l_2...,l_L}$; in this work, we assume the input to the neural network to be $l_0$ and the model prediction is $l_L$. The parameters of the model are then given by $ \theta = [\theta_{1}, \theta_{2}, ....\theta_{L}]$, where $\theta_{l}$ represents the parameters of model $M$ corresponding to the layer $l$. The first step in implementing a split learning design is to split the model into different parts so that the different parts can be implemented at different locations, i.e., in a model with 15 layers, the first 3 layers may primarily be computed in the end device which is generating or storing the input data, while the remaining layers may be computed in the edge device or the cloud. Similarly, for training cases, the backpropagation will be calculated at different locations as well, i.e., the last 12 layers may be computed in the edge/cloud for our example case, and then the needed backpropagation information may be sent to the end device for backpropagation calculation for the remaining 3 our of 15 layers. 

\begin{figure*}[t]
  \centering
  \resizebox{\linewidth}{!}{
    \includegraphics{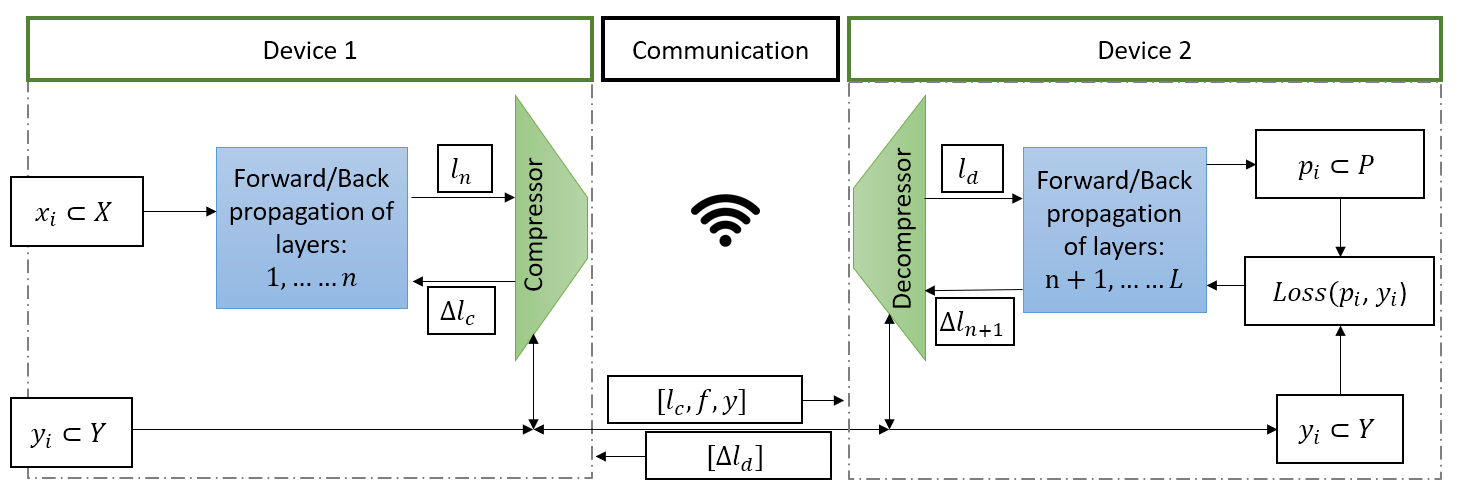}
  }
  \caption{System Overview: split learning representation across multiple devices with the compression-decompression module involved} 
  \label{fig:systemOverview}
\end{figure*}

Figure \ref{fig:systemOverview} shows an overview of our architecture for split implementation of a deep learning model. Let us take the model to have $L$ layers, where the splitting takes place after $n$ layers such that layers $1,...,n$ are computed at device 1, and the remaining $n+1,....L$ layers are computed at device 2. Any input $x_i \in X$ is generated at or provided to device 1, along with the corresponding real output $y_i \in Y$. In the forward propagation phase, the input $x_i$ is computed through the first $n$ layers in device 1 to give us the output of layer $n$, which is $l_n$. But since $l_n$ may be very large compared to the desired communication constraints, we introduce a compression mechanism, which we will call  compression module, to reduce the size of $l_n$ through a lossy compression method that compresses $l_n$ to $l_c$. This compression module is also computed on device 1. Once the data is communicated to device 2, a decompression module acts upon $l_c$ to give $l_d$, which becomes input for layer $n+1$ . Then the remaining $L-n$ layers are computed on device 2 to give the prediction output $p_i$.

As we note in figure \ref{fig:systemOverview}, the communication from device 1 to 2 not only carries with it $l_c$ but also two other variables, $f$ and $y$. $f$ is a filter parameter which we will discuss later while talking about compression and decompression modules (we will refer to them together as compression-decompression module.); $f$ is used to define which feature maps are communicated across the devices. While $y_i$ is the real output corresponding to the input $x_i$, $y_i$ is not needed in case of inference and we conclude after getting the prediction $p_i$, which may be sent back to device 1 and, in any case, it is not a communication costly task, but for training, next steps remain. Taking $p_i$ and $y_i$, we calculate the loss $\delta L(p_i,yi)$, which is then back-propagated through the neural network and decompression module to obtain $\delta l_d$. $\delta l_d$ is transmitted to device 1 where the remaining backpropagation can take place. Thus, both parts of the vertically split neural networks are trained in two different devices during the training phase.

While this unit may be a node in a multilayer perceptron or feature maps in CNNs, the method is designed to achieve compression for a range of general NNs; since the names are different in different context, we will refer to this unit as feature map for the sake of brevity in explanation. An equivalent unit would be the nodes at layer $n$ for multilayer perceptron. 

\begin{figure}[]
  \centerline{
    \includegraphics[width=0.40\textwidth]{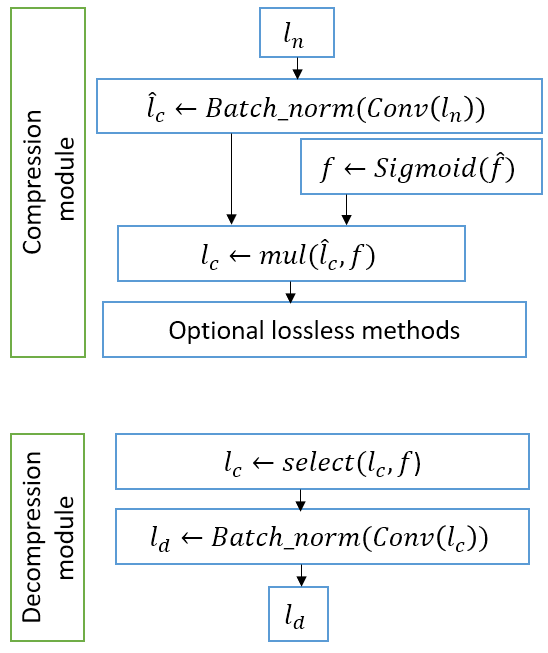}
  }
  \caption{Illustration of Compression and Decompression modules} 
  \label{fig:compModule}
\end{figure}

\textbf{Compression and decompression Modules:}  In this section, we will discuss the compression and decompression modules mentioned earlier. The first and optional step for the input to the compression module, $l_n$, is to go through an encoder mechanism to achieve a resolution compression for each node/feature map. This is an optional method and dependent on the kind of neural network being implemented. For multi-layer perceptron, the nodes are simply a real number and do not require this kind of compression. In a convolutional neural network for a potential resolution compression, this is achieved by controlling the kernel dimension and stride dimension of the convolution layer. We define resolution compression as the compression in size of each feature map of the input layer $l_n$, and resolution compression factor $r$ as the reduction in size of the feature map. For a given $r$, to achieve the desired compression, we define kernel height as $kc_h=2+r$, kernel width as $kc_w=2+r$ stride value as $sc=r$ and padding value as $pc=r$. Then the values are passed through a batch normalization layer. We will refer to this output as $\hat{l}_c$.  

Next we will explain the core compression and decompression modules as illustrated in figure \ref{fig:compModule}; we will also refer to them together as compression-decompression module. We begin as input $l_n$, the output of $n^{th}$ layer, with $\tilde{\phi}$ entries/nodes/feature maps (we will refer as feature map as usual). For the output of the compression module above, we obtain $\phi$ number of feature maps where $\phi \leq \tilde{\phi}$, and the compression filter $\hat{f}$ is a vector of size $\phi$. At the beginning of training we initialize $\hat{f}$ as a learning parameter of the neural network, with each value $\hat{f}_k \in \mathbb{R}$ where $k=1 \dots \phi$. After passing through a sigmoid to obtain $f$ from $\hat{f}$, we multiply each element of $f$ to the corresponding feature map, i.e, for layer output $\hat{l}_c$ with $\phi$ outputs, $\hat{l}_{c,k}$, the $k^{th}$ feature map, is now $l_{c,k}=f_k \cdot \hat{l}_{c,k}$. We will refer to this operation as "mul" for convenience, as referred in figure \ref{fig:compModule} as well, and the output is $l_c$. This method allows for further lossless compression methods, optionally. This compression module can be considered a method of compression similar to encoding but the level of compression is controlled by training the entries in the vector $f$ to either be close to zero or not. The nodes corresponding to zero value need not be used further / communicated further, and we will explain later how $f$ is trained to represent a desired compression level.   

Following the compression, the data is communicated over to device 2, where the decompression has to take place, with the decompression module restoring the resolution of the feature maps to what it was during $l_n$, and gives an output $l_d$. A communication budget $B$ is used to limit how many of the feature maps are communicated from device 1 to device 2, i.e., the feature map $l_{c,k}$ is communicated if and only if the corresponding $f_k$ is larger than $B$ entries in $f$. 

$l_d$ is then processed further with the remaining $n+1,...,L$ layers to produce the prediction $p$, to be compared against the real output $y$ during the training phase. So, during the training phase, we have two learning objectives: first is training the traditional learning parameters (for example, weights and biases) of the model, and second is learning the values of filter $f$ such that a communication budget $B$ is satisfied. In this work, we define a single loss function, that is used to train both sets of parameters, with the loss function $Loss(p,y)$ being an arbitrary loss function that is traditionally used to train a model, and a pruning loss function $pruneLoss(f,B)$, which is defined as:

\begin{equation}
    \label{eqn:pruneLoss}
    pruneLoss =
    exp \left( \delta \cdot \left[ \left( \sum\limits_{ {i \in 1,..,\phi} }{f_i}  \right) - B \right] \right) 
    + \lambda exp \left( - \delta \cdot \left[ \left( \sum\limits_{ {i \in 1,..,\phi} }{f_i}  \right) - B \right] \right)
\end{equation}

In equation \ref{eqn:pruneLoss}, the exponential functions act as a soft-constraint, or barrier functions, that let the budget be violated but at a great cost, and as the $pruneloss$ is minimized, the number of nodes selected can come in agreement with the given budget constraint $B$. There are two exponential functions, with the second one (negative exponential) ensuring that the budget constraint is not over-corrected; this is not necessary but improves the results. Note that $0 \leq \lambda \leq 1$.

The total loss $totalLoss$ is then defined as:

\begin{equation}
    \label{eqn:totalLoss}
    totalLoss = Loss + \epsilon \cdot pruneLoss
\end{equation}

In equation \ref{eqn:pruneLoss}, $\delta$ is a value defining how the violation of budget is handled as a soft constraint and in equation \ref{eqn:totalLoss}, $\epsilon$ is a value defining how the $pruneLoss$ is weighed against $Loss$ during the training.   

The method described above, including the selected approach of adding pruneloss (shown in equation \ref{eqn:pruneLoss}) to the loss, affords certain benefits to the process of training and implementing deep learning models in a networked system. 

\begin{itemize}

\item First, notice that the total loss function allows for both  $loss$ (associated with accuracy of the model) as well as the $pruneLoss$ (associated with the satisfaction of compression level) to be learned at the same time. We design and implement the method in such a way that the deep learning approach, i.e. implementation in the GPU, can be done for both kinds of learning. Hence these objectives can be trained in parallel. 

\item Second, notice that the parameter that controls the level of compression, $f$, is the only parameter in this deep learning architecture that needs to be controlled for different compression levels. So by training for one compression level, i.e., a particular value for $\sum_{ {i \in 1,\dots,\phi} }{f_i}$, at the beginning, and then changing this value according to budget without resetting other parameters, we are able to quickly retrain for different compression levels through this form of transfer learning. 

\item Third, by dynamically changing compression levels during learning, we are able to acquire different deep learning architectures suitable for different communication environments. 

\end{itemize}

As we discussed earlier, this is more of a skeleton method that the aforementioned techniques, "deprune" and "prune", employ to achieve their goals.

\subsection{Deprune Method}
\label{subsec:DePruning}

When the source of the data, i.e., device 1 (client device) in the formulation above and devices like IoT devices in the real world, are incapable of completing the learning process by themselves, or avoid learning in one location for the sake of saving on-device energy, split learning can be employed to share the task with the more powerful connected devices, i.e., device 2 (server) in the formulation above or devices like edge computers. In this instance, if the split learning were to proceed as is without any compression, depending on the model and the location where the splitting happens, a large amount of data would have to be transmitted across the network. For instance, with the convolutional neural networks, a huge number of feature maps would have to be transported across the network for split learning. A solution to this problem is to implement compression, and with lossy compression, we are able to further increase the compression as discussed earlier. However, there is a loss of accuracy for the model when too much compression is implemented;  the work around which lets us achieve both a lower training time and a higher accuracy level is the 'deprune' method based on our formulation.

In the 'deprune' method, we begin by taking a filter with a much smaller communication budget, i.e., allowed number of $1's$ in $f$, denoted by $b$ and guided by $B$, is much smaller than size of $f$ given by $\phi$. For instance, after certain layers of computation, at layer $n$, a VGG neural network has $128$ feature maps, but we may want to begin the training by only allowing $4$ feature maps to be communicated from device 1 to device 2. After training the neural network for a few epochs, we would then increase the number of feature maps from $b<<\phi$ to maximum value of $\phi$, i.e., in case of the example VGG, by increasing the allowed number of features maps to be communicated, $b=4$, to $b=128$. This method is elaborated in algorithms  \ref{algorithm:dePrune1} and \ref{algorithm:dePrune2}. 

Algorithm \ref{algorithm:dePrune1} takes place in the client device, or as we referred earlier, device 1. For a model $M$ with training hyper-parameters $H$, our goal will be to train the parameters corresponding to the layers $\{1,....,n\}$, with the help of two lists: $Budgets$, where each entry gives the budget $B$ (as defined earlier) under which the compression is supposed to operate at a given instance of training, and $Epochs$ where each entry corresponds to the number of epochs to be trained for a given budget $B$. The end goal here is to obtain the trained parameters in client device given by $\theta_{client}$. Please note that this value consists of training parameters for first $n$ layers as well as the parameters for compression module, i.e., $f$. The data is obtained for certain batch sizes as defined by $H$ through a batch-loading mechanism (any sort of) in the form of $(X_t,y_t)$, and then we perform a forward pass on input $X_t$ through the first $n$ layers to obtain the output $l_n$. 

\begin{minipage}{0.46\textwidth}
\begin{algorithm}[H]
    \caption{dePrune training at client device}
    \label{algorithm:dePrune1}
    \textbf{Input:}  $M$, $H$, $n$, $Budgets$, $Epochs$, $l_k$, $\gamma$\\
    \textbf{Output:} Trained parameters $\theta_{client}$\\
    \begin{algorithmic}[1]
    \STATE Initialize:  Parameters $\theta _1$ for $M's$ layers $\{1,....,n\}$\\

    \FOR{$B \in Budgets$}
        \FOR{$i \in Epochs[B]$}
          \STATE $selectLr(l_k, \gamma)$
            \FOR{$X_t,y_t \in batchLoader$}
                \STATE $l_n \gets ForwardPass(X_t)$
                \STATE $l_c,f \gets CompressionModule(l_n)$
                \STATE $Send(l_c,y_t,f)$
                \STATE
                \STATE $Receive(\Delta l_d)$
                \STATE $Backprogagation(\Delta l_d)$
            \ENDFOR
        \ENDFOR
    \ENDFOR
    \end{algorithmic}
\end{algorithm}
\end{minipage}
\hfill
\begin{minipage}{0.48\textwidth}
\begin{algorithm}[H]
    \caption{dePrune training at server device}
    \label{algorithm:dePrune2}
    \textbf{Input:}  $M$, $n$, $Budgets$, $Epochs$, $\delta$, $\epsilon$, $l_k$, $\gamma$\\
    \textbf{Output:} Trained parameters $\theta_{server}$\\
    \begin{algorithmic}[1]
    \STATE Initialize:  Parameters $\theta_2$ for $M$ for $M's$ layers $\{n+1,....\}$\\
    \FOR{$B \in Budgets$}
        \FOR{$i \in Epochs[B]$}
         \STATE $selectLr(l_k, \gamma)$
            \WHILE{running()}    
                \STATE $Await(l_c,y,f)$
                \STATE $l_d \gets decompressionModule(l_c,f)$
                \STATE $p \gets ForwardPass(l_d)$
                \STATE $\Delta l_L \gets totalLoss(p,y_t)$ given by equation \ref{eqn:totalLoss}
                \STATE $\Delta l_d \gets Backpropagation(l_L)$
                \STATE $Send(\Delta l_d)$
             \ENDWHILE
        \ENDFOR
    \ENDFOR
    \end{algorithmic}
\end{algorithm}
\end{minipage}

\hfill
\hfill

Next step is to perform compression on $l_n$ using the compression module as discussed during the formulation section and illustrated in figure \ref{fig:compModule}, which gives as output the parameters $l_c$ (output of compression) and $f$ (the filter). Then  $(l_c,y,f)$ is sent to the server device by the client device. 

Once $(l_c,y,f)$ is received by the server device, which is awaiting a request from client as shown in algorithm \ref{algorithm:dePrune2}, $l_c$ is decompressed using the decompression module as explained in the formulation section \ref{subsec:formulation}; output $l_d$ then goes through forward pass for the remaining layers to obtain the prediction $p$. Using equation \ref{eqn:totalLoss}, we compute the error for the given forward pass, and then the obtained $\Delta l_L$ is back-propagated through $L,...,n+1$ layers as well as the decompression module. Then the output of backpropagation on server device may be sent to the client as $\Delta l_d$. Client device, which is until now waiting to receive $\Delta l_d$ will now use this value from the server device to finish the training step.

An important function that is yet to be explained for the deprune methods as shown in algorithms \ref{algorithm:dePrune1} and \ref{algorithm:dePrune2} is the dynamic selection of learning rate ($selectLr(l_k)$). During $selectLr(l_k)$ for each non-initial budget level, i.e., when we transfer to higher budget level, the learning rate is set to a higher value by a factor of $\gamma$ for the first $l_k$ epochs. This approach was observed to significantly help avoid unwanted local minima during training and help achieve higher accuracy as we moved to higher budget levels.

\subsection{Prune Method}
\label{subsec:Pruning}

When the source of the data, i.e., device 1 (client device) from the formulation section and devices like XR devices in the real world, are not capable of completing the inference process in a timely manner or of doing so in an energy-permissible way, split learning can be employed to share the task with the more powerful devices in the network, i.e., device 2 (server) in the formulation section or devices like edge computers. In such an instance, if the split learning were to proceed without any compression, depending on the model and other requirements like latency constraints for inference, it might not be feasible to complete the task with just the end device running the inference model. Hence, we could again rely on compression, including lossy compression to achieve network performance (such as lower latency and network usage) that would otherwise be impossible when relying only on a weak end device and/or split learning without compression. 

As we implement more compression, we expect the model accuracy to fall, as less information can be transmitted from the source device / client device to the server for further inference. In any case, for different levels of compression, we will end up achieving different levels of model efficiency \cite{eshratifar2019bottlenet}. However, depending on latency/ network usage requirements, the client may want to opt for a different level of this accuracy-compression trade-off. While the lossy compression model is shown to be efficient, it does take quite a bit of time to train for a given compression level as training needs to start again from the beginning for that compression level. Therefore, if we want to train for different compression levels, the completion of training for each process is anticipated to require a substantial amount of time. With the help of our method, 'prune', based on the deep learning architecture as defined in the formulation section, we are able to implement transfer learning approach and generate multiple approximate models with similar accuracy (or, slightly reduce accuracy), but significantly reduced network costs and computation costs at the client devices.

In the pruning method, we begin by taking a filter with the largest possible filter size, i.e., where the communication budget is maximum, i.e., all the feature maps may be communicated so $B=len(f)/\phi$. For this particular model, we allow the training to fully proceed, and then let this be the "base model" upon which transfer learning may take place. Let $Budgets$ be a vector containing each desired feature map compression levels with $k^{th}$ compression level given by $B_k$. Then the model corresponding to $k^{th}$ compression level is given by $M_k$. Hence, after the full model with no compression ($B=\phi$) is trained to obtained parameters for $M_0$, we do the same for different $k$ compression levels as defined in $Budgets$ using the transfer learning approach as defined in algorithm \ref{algorithm:prune1}, which is designed to give us a set of trained parameters $\Theta$ where each element $\theta_k$ is parameters for model $M_b$ corresponding to compression level $B_k$.   

Only for budget $B = max(Budgets)$ where the feature map compression is not implemented, the set of parameters $\theta_b$ is randomly initialized as if training a new deep learning architecture; for every other compression level, we transfer the parameters learned from previously trained model. Then for a certain epochs given by $Epochs[B]$, we train the deep learning architecture. Later, while discussing results, we will show that $Epochs[B]$ only needs to be larger when $B=max(Budgets)$ but can be much smaller for the remaining compression levels. This way, the transfer learning approach can help us train many subsequent approximate models quickly. Notice that the models are still split into two parts, with one side computing the first $n$ layers, i.e. ${1,...,n}$ along with a compression module, and the other side computing the remaining $L-n$ layers after going through decompression module. 

This still is in keeping with our split learning formulation, but the major difference here when compared against the deprune method is that the training phase need not be run in different devices, but may be run in one powerful device that trains the different deep learning architectures with different network cost and performance trade-offs. Once the training is completed, the developed models represented by output of the prune method/algorithm \ref{algorithm:prune1}, $\Theta$, may be used by the client and server devices. In that case, the inference method is equivalent to the forward pass in the client device (as shown in algorithm \ref{algorithm:dePrune1}) followed by the forward pass in the server device (as shown in algorithm \ref{algorithm:dePrune2}). The client would in that case only expect the prediction $p$ from the server device; needless to say the learning parts including loss predictions are not necessary for these inferences. 

One important step after completing the training for each budget level is the $resetprune$ shown in algorithm \ref{algorithm:prune1}, where the filter parameters are reset. Notice that $f$ is always obtained by passing through sigmoid, and after training at certain budget level $B_{k1}$, we then move to budget level $B_{k2}<B_{k1}$. So here by making sure that the entries in $f$ that were already greater than $1$ during training for $B_{k1}$ are randomized during the training for $B_{k2}$, we make sure that those filters can potentially be trained to approach $0$, or else further compression would not be likely. An additional noteworthy point is that the pruning algorithm can be executed with various levels of resolution compression during CNN training. As was the case for the deprune method, we can also employ the dynamic selection of learning rate ($selectLr(l_k)$). During $selectLr(l_k)$ for each non-initial budget level, i.e., when we transfer to lower budget level, the learning rate is set to a higher value by a factor of $\gamma$ for the first $l_k$ epochs.

Algorithm \ref{algorithm:prune1} summarizes our prune method as an algorithm as discussed in section \ref{subsec:Pruning}.

\begin{algorithm}[]
    \caption{prune training}
    \label{algorithm:prune1}
    \textbf{Input:}  $M$, $H$, $n$, $Budgets$, $Epochs$, $l_k$, $\gamma$\\
    \textbf{Output:} Trained parameters set $\Theta$\\
    \begin{algorithmic}[1]

    \FOR{$B \in Budgets$}
        \IF{$B == max(Budgets)$}
            \STATE Initialize: Randomized parameters $\theta_B$ for $M_0$\\
            \STATE Initialize $\Theta \gets \{ \}$
        \ELSE
            \STATE Transfer: $\theta_B \gets \theta$
        \ENDIF
        
        \FOR{$i \in Epochs[B]$}

            \FOR{$X_t,y_t \in batchLoader$}
                \STATE $l_n \gets ForwardPass[l_1,...l_n](X_t)$
                \STATE $l_c,f \gets CompressionModule(l_n)$
                \STATE $l_d \gets decompressionModule(l_c,f)$
                \STATE $P \gets ForwardPass[l_{n+1},...l_L](l_d)$
                \STATE $\Delta l_L \gets totalLoss(P,y_t)$ given by equation \ref{eqn:totalLoss}
                \STATE $\Delta l_d \gets Backpropagation(l_L)$
                \STATE $Backprogagation(\Delta l_d)$
            \ENDFOR
        \ENDFOR
        \STATE $\theta \gets \theta_B$
        \STATE $\Theta[M_b] \gets \theta_B $
        \STATE $resetprune$
    \ENDFOR
    \end{algorithmic}
\end{algorithm}


\section{Evaluation}
\label{sec:eval}

In this section, we will show the results demonstrating the efficiency of our method at creating the deep learning architectures in network system paradigms, with a focus on overcoming network costs, i.e., reduced network usage and decreased communication latency. But we will also focus on overall processing time. In subsection \ref{subsec:evaldePrune}, we will evaluate the depruning method, alongside its implementation on testbeds. And in subsection \ref{subsec:evalprune}, we will present and describe the benefits of the pruning method.

\subsection{Results: dePrune Method}
\label{subsec:evaldePrune}

A primary goal of our deep learning architecture, and of the deprune method, is to reduce bandwidth usage and improve latency during the training process; it is then by extension that the benefits are afforded to the end device in the form of manageable reduction in computation cost/energy consumption through split learning. We begin by showing that our method is able to significantly reduce bandwidth consumption on a realistic testbed, while also being capable of improving learning time/latency. Then we run simulations over python implementations to demonstrate the reduction in the network usage as we employ the deprune method on different models, datasets, split locations and resolution compressions. The code developed for these simulations are open-sourced. \footnote{\url{https://github.com/amudvari/VCCprune/tree/artifacts}}

In order to show that our method allows training to continue with a reduction in bandwidth usage over time, we show that less bytes have to be communicated over the training period (to reach the same level of accuracy in the same number of epochs) with our method added to a distributed/split learning paradigm. Additionally, in order to show that the end-to-end latency of learning episodes can be reduced on average, we show that more samples are processed using our method over time, when a device has to offload certain fraction of the deep learning task to a server. Notice that offloading raw data (i.e., image) can be considered a special case where offloading takes place before layer $1$ as opposed to certain arbitrary layer $n+1$.


The testbed consists of an embedded GPU serving offload requests, and a laptop hosting data sources which offload
processing. The embedded GPU uses CUDA for most of the tensor processing, and CPU for the rest of the computations,
including network message serialization and de-serialization. The laptop uses CPU for all of the computations,
including tensor processing. The laptop uses a commodity CPU with 8 cores and a base clock frequency of 1.60 GHz.
The processing speed of the GPU according to the manufacturer's specification is 22 TOPS (INT8 precision).
The laptop and the embedded GPU communicate over a physical Wi-Fi 6 channel.

The software framework used in benchmarks is open-sourced
\footnote{\url{https://github.com/AnteronGitHub/sparse}}. In the
benchmarks, the laptop runs multiple parallel data source processes, each of which share the same wireless channel. The
processes were containerized, and their deployment was automated by using Kubernetes, which allows for much better orchestration and scaling among other benefits. To emulate the constrained
processing power of mobile devices, and to avoid physical hardware saturation from other applications running in the
laptop affecting the measurements, each of the data sources' CPU allocation was limited virtually. Furthermore, the
number of data source processes running in the laptop was set to eight, which the laptop had enough RAM to run without
having to use swap memory.

In the benchmarks, we measure the task throughput in terms of samples processed, as well as the data traffic in terms
of bytes sent over the wireless channel. The tasks throughput is measured per data source, while the data traffic in
the wireless channel is the sum of all of the data sources' traffic. The benchmark measurement starts after processing
the first batch in order to avoid the cold start resulting from CUDA JIT compilation.

We run benchmarks with eight data sources, limiting the CPU usage of each data source to 40\% cores. The resource usage
of the embedded GPU was not limited. Each data source processed samples in batches of 16 samples, and processed 32
batches in 4 epochs. Split training without pruning does not include the additional neural network layers, and the
compression/de-compression modules, resulting in less overall processing. 

\begin{figure}[!tbp]
  \centering
  \begin{minipage}[b]{0.45\textwidth}
    \includegraphics[width=\textwidth]{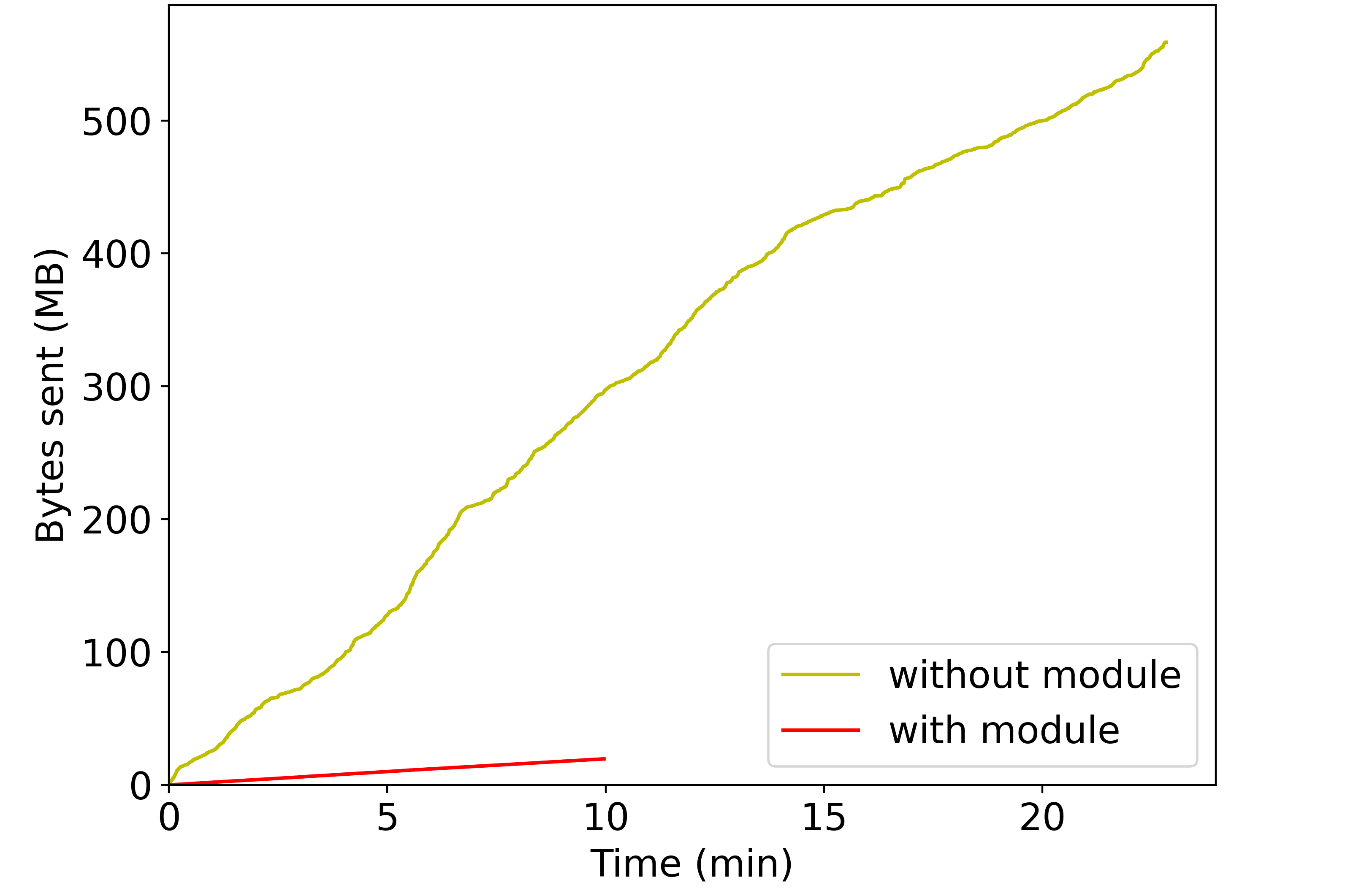}
    \caption{Data traffic in the Wi-Fi channel during the benchmark execution.}
    \label{fig:benchmark-data-traffic}
  \end{minipage}
  \hfill
  \begin{minipage}[b]{0.45\textwidth}
    \includegraphics[width=\textwidth]{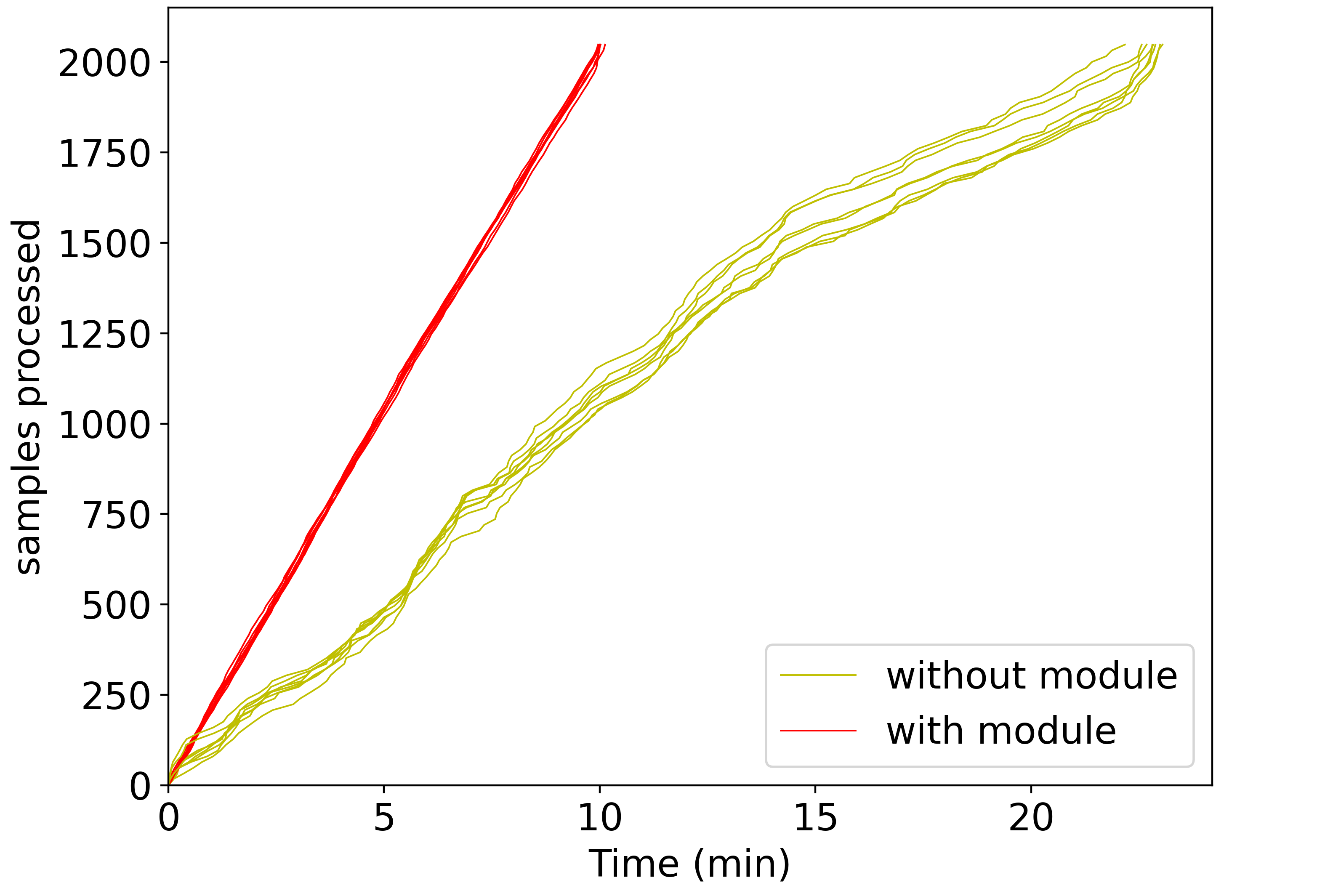}
  \caption{Throughput for each data source during the benchmark execution. The plots are interpolated linearly to make them more readable.}
  \label{fig:benchmark-throughput}
  \end{minipage}
\end{figure}

We run the experiment for CIFAR10 dataset \cite{krizhevsky2009learning} running VGG11 model \cite{simonyan2014very} with our compression-decompression module (referred as 'with module') and compare against implementation that does the split learning without any compression-decompression module \cite{kang2017neurosurgeon} (referred as 'without module); in both cases splitting takes place at 5th layer with the compression module having $f=4$ in case of our method. These layer selections are arbitrary and later we show that our method works across different layers. As seen in figure \ref{fig:benchmark-data-traffic}, the network usage is reduced significantly with our method when compared against split learning (labelled 'without module'), i.e., the improvement is about 25 times less network usage; and the training throughput as shown in figure \ref{fig:benchmark-throughput} is reduced by about 2.5 times. The goal here was to demonstrate that the deployment of our method in a wireless environment can efficiently reduce the network usage, and increase task throughput, i.e., provide improved total latency for the completion of the learning task. 

\begin{figure}[]
  \centering
  \begin{minipage}[]{0.45\textwidth}
    \includegraphics[width=\textwidth]{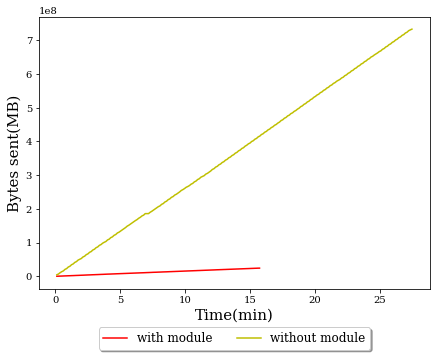}
    \caption{Data traffic in the Ethernet channel during the benchmark execution.}
    \label{fig:gpus_benchmark-data-traffic}
  \end{minipage}
  \hfill
  \begin{minipage}[]{0.45\textwidth}
    \includegraphics[width=\textwidth]{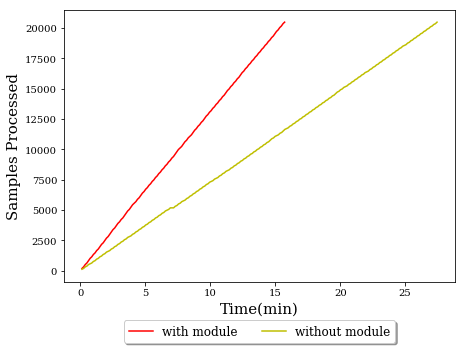}
  \caption{Throughput for each data source during the benchmark execution.}
  \label{fig:gpus_benchmark-throughput}
  \end{minipage}
\end{figure}

In the next case, we run the experiments with a single data source running CIFAR10 dataset and learning on VGG11 model, with a batch size of 64, and other similar settings as compared with testbed results in section \ref{subsec:evaldePrune}. The main difference here is the hardware being used, with both the client and the server devices being CUDA-enabled nodes and connected to the network over Ethernet. Specifically, the client and the server nodes are set up with an NVIDIA GeForce RTX 3090 and an NVIDIA GeForce RTX 4090 respectively, and the Round Trip Time between these two nodes over Ethernet connection is {$0.670\pm0.219ms$}. As shown in figure \ref{fig:gpus_benchmark-data-traffic}, our method ('with module') is able to reduce the total network usage over split learning without our module ('without module') by 30x, while as shown in figure \ref{fig:gpus_benchmark-throughput}, the total throughput is reduced by 1.7 times. Of course, this throughput performance was a case where computation was the bottleneck and not the network, since Ethernet was used; when network is a bottleneck, it is shown that the method is more useful; i.e., when WiFi or cellular network is involved, or when network is congested, the method becomes more useful. Of course, this throughput performance was a case where computation was the bottleneck and not the network, since Ethernet was used; when network is a bottleneck, it is shown that the method is more useful; i.e., when WiFi or cellular network is involved, or when network is congested, the method becomes more useful.

In figures \ref{fig:benchmark-data-traffic} or \ref{fig:gpus_benchmark-data-traffic}, the cumulative bytes of network traffic used up can be interpreted in two ways: the instantaneous slope can be used to see that the bandwidth consumption was much lower for our method most of the times, signifying that the bandwidth consumption (bytes sent per unit time) was lower. The red line (our method, 'with module') ends much faster, signifying that for same level of accuracy and same number of epochs run, the training session was finished faster. Similarly, the slope of the lines in figures \ref{fig:benchmark-throughput} and \ref{fig:gpus_benchmark-throughput}, show that our method (red lines, 'with module'), have a higher slope, denoting faster completion of each training sample, denoting a better end-to-end processing latency for the training steps. These experiments have shown that the method could work in different network and hardware settings. Furthermore, the use of multiple workers in case of WiFi network also demonstrates the usefulness of this method for a scaled deployment (with multiple users participating in the learning environment). 

Next, we show how the method runs with less bandwidth consumption, while reaching the same level of accuracy as non-compressed offloading of the tasks to servers, across different configurations.  

\begin{figure}[!tbp]
  \centering
  \begin{minipage}[b]{0.42\textwidth}
    \includegraphics[width=\textwidth]{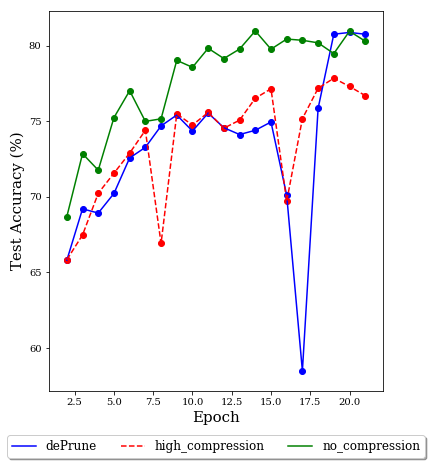}
    \caption{Progression of training under deprune method, cifar10 dataset} 
    \label{fig:compareModuleCF}
  \end{minipage}
  \hfill
  \begin{minipage}[b]{0.42\textwidth}
    \includegraphics[width=\textwidth]{{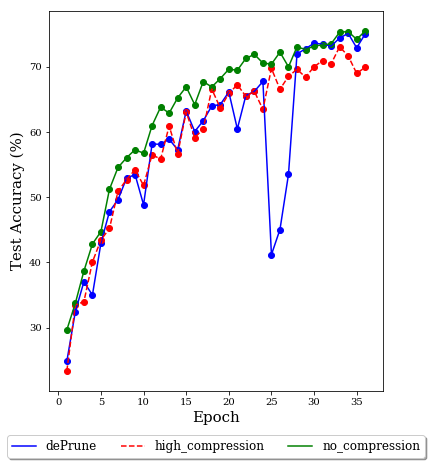}}
  \caption{Progression of training under deprune method, imagenet dataset} 
  \label{fig:compareModuleIN}
  \end{minipage}
\end{figure}

We ran simulations to show the network usage improvement under different circumstances. In figure \ref{fig:compareModuleCF} we train a VGG11 \cite{simonyan2014very} NN with CIFAR10 dataset \cite{krizhevsky2009learning} using 3 different approaches. 'no-Compression' method refers to a process in which the filter size $f$ as defined during the formulation is maximum, which means all the data/feature maps are sent from client device to server device. When we consider the layer to be $0_{th}$/before first layer, it represents simply offloading the raw data to the server (i.e., cloud).  On the other hand, 'high-compression' refers to the case where the compression is implemented so that very few number of feature maps are allowed to be communicated. Finally, 'deprune' is our method as described earlier in subsection \ref{subsec:DePruning}. 'deprune' is a method where for a much greater part of the training epochs, we will only train under the lower compression level / lower value of $f$, where much fewer filters are communicated across the network. In case of figure \ref{fig:compareModuleCF}, we will train with the filter size $f=4$ out of $128$ for the first 15 epochs, and then we may train with full feature size $f=128$ for the remaining time. Here we see that the level of accuracy towards the end is similar to training with full budget, given by test accuracy in percentage in figure \ref{fig:compareModuleCF}, reached by epoch 18. High compression option here is with $f=4$ and no compression option always has $f=128$, i.e., full budget. Here, we realized that while the accuracy level achieved by 'deprune' is similar to that of the accuracy achieved by 'no-compression' approach by 18 epochs, the improvement in network utilization would be by 420 percent. The accuracy is improved by 4 percent as compared to high compression case by then. In case of this illustration figure \ref{fig:compareModuleCF}, the hyperparameters we used were: Stochastic Gradient Descent (SGD) as optimizer, learning rate $l$ of 1e-5, weight decay $\gamma$ of 5e-4, and split layer $n=5$.

Next, we tested the experiment for a different dataset, Imagenet100, which is a subset of ImageNet-1k dataset\cite{russakovsky2015imagenet} that contains 100 out of the 1000 classes in the original dataset, using the same approach as earlier in case of CIFAR10 dataset, but for images with higher resolution. The result is illustrated in \ref{fig:compareModuleIN}, where comparison is once again made with no-compression and high-compression cases. Here again, training is done for a lower value of $f=4$ for 25 epochs, before switching to $f=128$ for the remaining of the training. We see that by epoch 30, the accuracy of 'deprune' method is already similar to the accuracy when no compression happens. Without a loss of obtained accuracy, within these 30 epochs, the network usage is improved by 418 percent. While by then the accuracy has improved by 4 percent as compared against high-compression case. In case of this illustration figure \ref{fig:compareModuleIN}, the hyperparameters we used were: Stochastic Gradient Descent (SGD) as optimizer, learning rate $l$ of 1e-5, weight decay $\gamma$ of 5e-4, and split layer $n=5$.

\begin{figure*}[]
  \centering
  \resizebox{\linewidth}{!}{
    \includegraphics{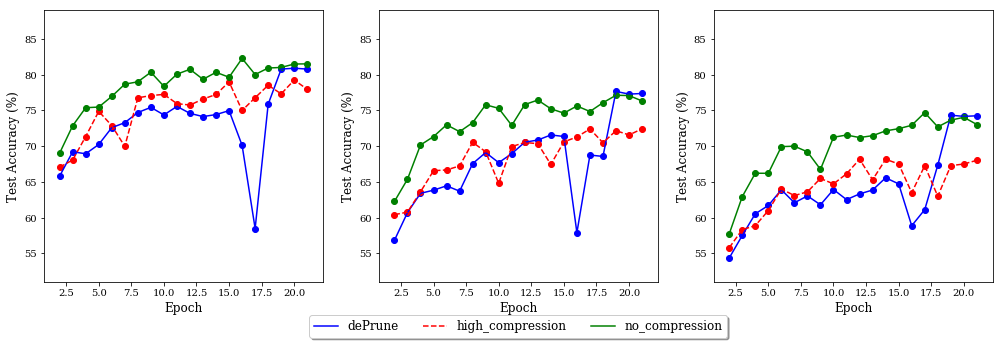}
  }
  \caption{deprune with different resolution compression factors. Left: no resolution compression, Center: by 2x2, Right: by 3x3} 
  \label{fig:depruneResComps}
\end{figure*}

We also repeated this same experiment with different resolution compression levels for CIFAR10 dataset, as shown in figure \ref{fig:depruneResComps}, which shows that these compression mechanisms work for different levels of resolution compression. As our algorithm uses a convolution method during compression, by increasing the convolution kernel for compression from 1x1 to 2x2 (center plot of \ref{fig:depruneResComps}) and by increasing to 3x3 (left plot of figure \ref{fig:depruneResComps}), we observe that our method of compression can work complementary to resolution compression methods. As expected in each case, as the compression factor increases from 1x1 to 2x2 and 3x3, the accuracy falls for each case, but in each case our method efficiently matches the maximum test accuracy (no-compression case) for given resolution compression level.     

\begin{figure*}[]
  \centering
  \resizebox{\linewidth}{!}{
    \includegraphics{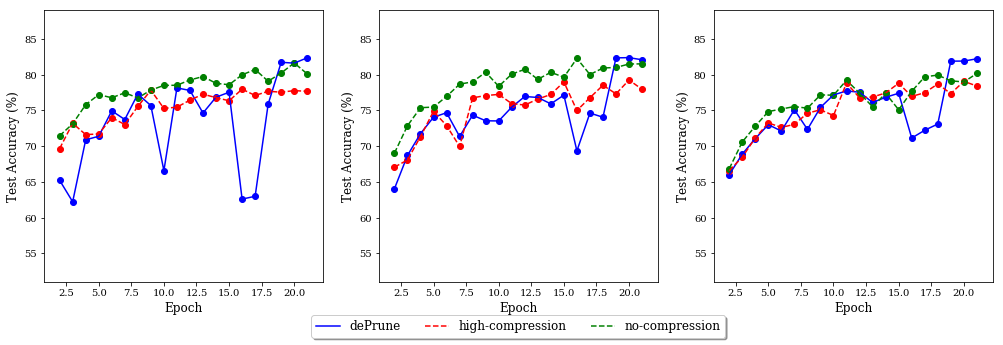}
  }
  \caption{Deprune with compression-decompression module after different layers. Left: before layer 1/Image, Center: after layer 6, Right: alter layer 9} 
  \label{fig:depruneDiffLayers}
\end{figure*}

As shown in figure \ref{fig:compareModule}, appendix \ref{subsec:appendixAddModule}, the addition of the compression-decompression module did not have a significant impact on the performance of the deep learning model, and the addition of this module at any location within the model did not change the behavior in any significant way either; the test accuracy was similar after any number of epochs for each of the cases. To make sure that the addition of compression-decompression module after any arbitrary layer results in similar efficient behavior, we ran the experiments again for CIFAR10 dataset with different locations of the split layer (in addition to original experiment where splitting occurs at layer 5), as shown in figure \ref{fig:depruneDiffLayers}. In cases where the compression-decompression module were kept at the 0th layer (i.e. compressing the input itself) as shown in figure \ref{fig:depruneDiffLayers} (left), after 6th layer as shown in figure \ref{fig:depruneDiffLayers} (center), and after 9th layer as shown in figure \ref{fig:depruneDiffLayers}(right), the efficient behavior from our deprune method as explained earlier was always observed. This effectively proves that the module could be added anywhere, as well as that the 'deprune' method can be run in any of these cases.

This shows that the method can be deployed in different ways and for different data, so that network resources are saved while training deep learning models in a split-learning environment.

\subsection{Results: Prune Method}
\label{subsec:evalprune}

As discussed earlier in the related works and methods sections, compression aware training, where we train deep learning models to split the tasks in a network and latency-efficient way, requires substantial training time. This is because we need to train the model for every given compression level. Our solution based on the compression-decompression module, 'prune', is able to generate models with different compression-latency trade-off by training the largest model first and then using a transfer learning approach to quickly generate approximate models, as described in subsection \ref{subsec:Pruning}. In this subsection, we will demonstrate the results of our method in reducing the training times for these approximate models in an efficient way.

\begin{figure*}[t]
  \centering
  \resizebox{0.9\linewidth}{!}{
    \includegraphics[width=\textwidth]{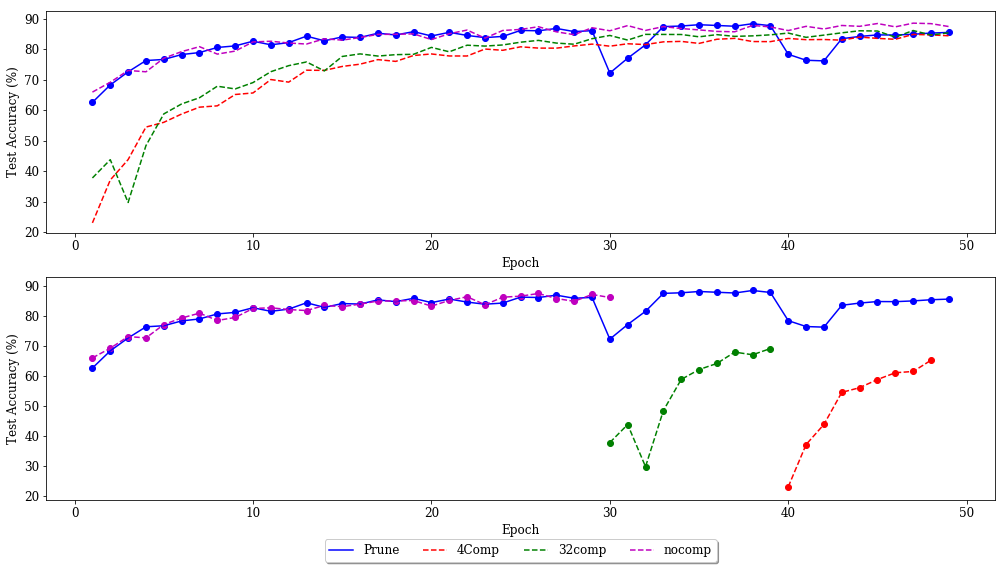}
  }
  \caption{Progression of training under prune method, CIFAR dataset} 
  \label{fig:prune_cifar10progress}
\end{figure*}

\begin{figure*}[t]
  \centering
  \resizebox{0.9\linewidth}{!}{
    \includegraphics[width=\textwidth]{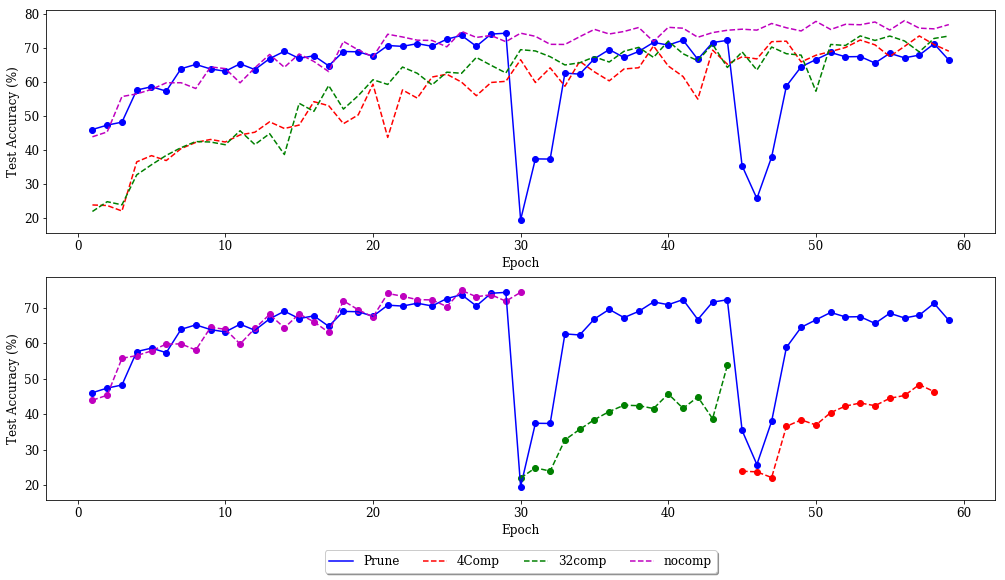}
  }
  \caption{Progression of training under prune method, STL10 dataset} 
  \label{fig:prune_stl10progress}
\end{figure*}

In figure \ref{fig:prune_cifar10progress}, we train a VGG11 \cite{simonyan2014very} neural network with CIFAR10 dataset \cite{krizhevsky2009learning} using our method, 'prune'. In this case, we begin by training the model first with no compression, i.e., filter size $f=128$ so all 128 feature maps in the layer are to be sent across network for the first 30 epochs. Then we reduce the number of filters to $f=32$, and then to $f=4$, training for each $f$ for a few epochs. In figure \ref{fig:prune_cifar10progress}(top), we compare our method Prune with situations where compression levels are no compression (labelled 'nocomp'), $f=32$ (labelled '32comp', representing a compression by 4 times), and $f=4$ (labelled '4comp', representing a compression by 32 times). The top image, figure \ref{fig:prune_cifar10progress}(top), is showing that our method could generate results that are comparable to the results generated by 'training from start', with compression by 4 times called 32 comp, compression by 32 times labelled 4comp, and no compression-case labelled nocomp. In these cases, which we call 'training from scratch' compression level is predefined and the model is trained without transferring any learned information from any other model, for each compression level. For our approach prune, training happens in epochs 0-30 for $f=128$, epochs 30-40 for $f=32$, and epochs 40-50 for $f=4$, and in all of those cases, accuracy to the 'training from scratch' approach is matched. 

In figure \ref{fig:prune_cifar10progress}(bottom), the bottom figure, we are able to see that after training for $f=128$ and transferring to $f=32$, the training time is significantly reduced, i.e. the accuracy after training for just 5 epochs has reached the maximum, while for 'training from scratch', this value is only achieved (inferred from figure \ref{fig:prune_cifar10progress}) after about 30 epochs; after 5 epochs accuracy is only 70 percent of maximum for the 'training from scratch' approach. Hence, this is an improvement of 6 times better training speed. Similarly, for the compression level of $f=4$, the training time was an improvement of about 6 times as well. In case of this illustration figure \ref{fig:prune_cifar10progress}, the hyperparameters we used were: Stochastic Gradient Descent (SGD) as optimizer, learning rate $l$ of 1e-5, weight decay $\gamma$ of 5e-4, and split layer $n=5$.

We repeated this experiment with STL10\cite{coates2011analysis}, as illustrated in figure \ref{fig:prune_stl10progress}. The same method as described in case of CIFAR10 above was implemented again. This time for our approach prune, training happens in epochs 0-30 for $f=128$, epochs 30-45 for $f=32$, and epochs 45-60 for $f=4$ as evident in figure \ref{fig:prune_stl10progress} bottom or top. As seen in \ref{fig:prune_stl10progress}(top), our method was able to match performance of 'training from scratch' with 4 times compression (labelled '32Comp' in figure \ref{fig:prune_stl10progress})  by epoch 7 as opposed to taking about 30 epochs to reach the maximum performance, which is a improvement of 4 times. As seen in \ref{fig:prune_stl10progress}(bottom), STL10 was even much slower than the CIFAR10 case of 'training from scratch'; here by epoch 7, in either cases of compression, 4 times compression with $f=32$ or 32 times compression with $f=4$, within 7 epochs, the training in case of 'training from scratch' had only reached about 60 percent of our method. In case of this illustration figure \ref{fig:prune_stl10progress}, the hyperparameters we used were: Stochastic Gradient Descent (SGD) as optimizer, learning rate $l$ of 1e-5, weight decay $\gamma$ of 5e-4, and split layer $n=5$. 

In Appendix \ref{subsec:appendixpruneResults}, we discuss further configurations for trying out the prune methods. So this section shows that we are able to obtain network-efficient deep learning architectures quickly. The next straightforward step in deploying our models, after training, would be to have them in edge-cloud split environment for quick inference, while giving the network management entity the choice of selecting between accuracy and latency; this could help make decisions to optimize service quality, fulfill latency constraint, or help with any other utility maximization. 

\section{Conclusion}
\label{sec:concl}

In this work, we developed a novel approach to training deep learning models that improve the network usage and the training latency. We designed a compression-decompression module, that when combined with our deprune algorithm, is able to significantly reduce network usage and make learning faster without significantly effecting accuracy; network usage for learning was improved by 4x when compared with split learning approach, and the accuracy was improved by 4 percent when compared with compression-aware split learning. Such algorithms will be very beneficial to IoT and other resource-constrained end devices in the next generation network architectures that must provision AI tasks. We also combined the compression-decompression module with our prune method to significantly reduce the training time for deep learning models that trade little accuracy for significant reduction in network consumption; here we showed that the training time could be improved by up to 6 times when compared with compression-aware split learning. Such methods could help develop orchestration options for AI task like AR/XR in the network constrained environments.

\begin{acks}

This research was in part supported by the Academy of Finland (grant number 345008), and the National Science Foundation CNS AI Institute (grant number 2112562), as well as the NSF-AoF FAIN project (grant number 2132573).

\end{acks}

\bibliographystyle{ACM-Reference-Format}
\bibliography{sample-base}

\section*{Appendix}
\appendix

\section{Effect of adding compression-decompression module to a neural network}
\label{subsec:appendixAddModule}

\begin{figure}[]
  \centerline{
    \includegraphics[width=0.5\textwidth]{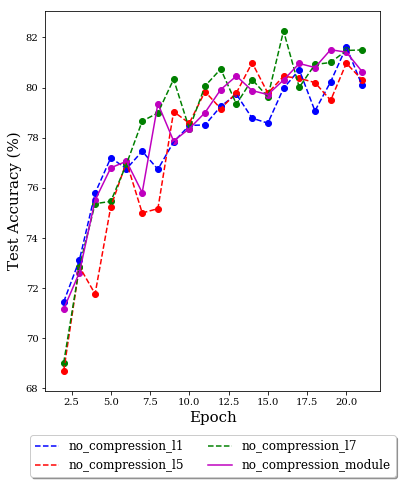}
  }
  \caption{Progression of training for models with compression-decompression modules after different layers along with a model without the module} 
  \label{fig:compareModule}
\end{figure}

In figure \ref{fig:compareModule}, we have shown that putting the compression-decompression module after different layers produces very similar learning behavior to not putting them anywhere, i.e., the original model. In other words, the learning trajectory and the outcome after training are very similar. The results are demonstrated after putting the module after no layers (i.e, image), after 5th layer and after 7th layer.




        


\section{Extended evaluation for prune method}
\label{subsec:appendixpruneResults}

\begin{figure}[]
  \centerline{
    \includegraphics[width=0.9\textwidth]{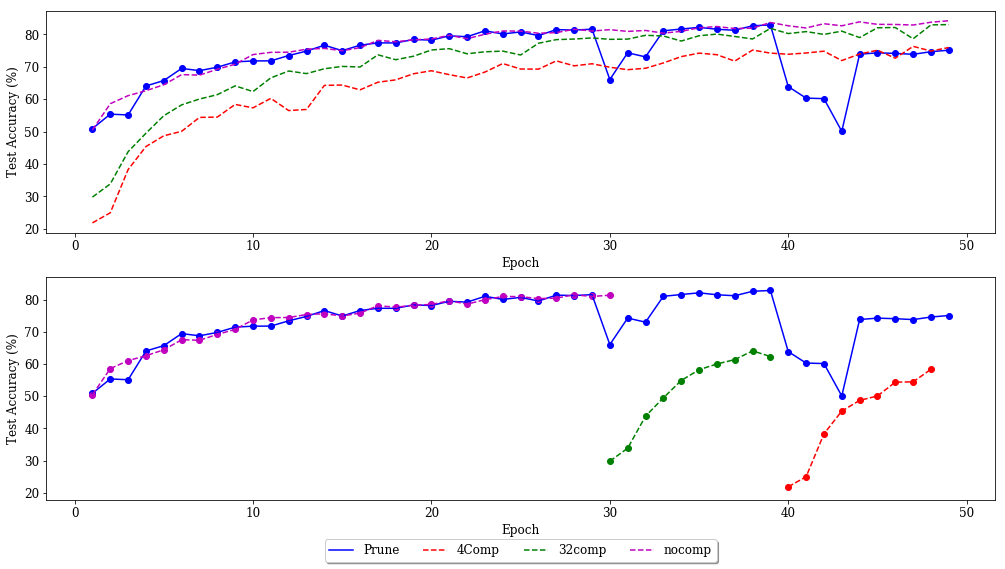}
  }
  \caption{Progression of training under prune method, for CIFAR10 dataset with resolution compression factor of 3x3} 
  \label{fig:prune_cifar10r3}
\end{figure}

\begin{figure}[]
  \centerline{
    \includegraphics[width=0.9\textwidth]{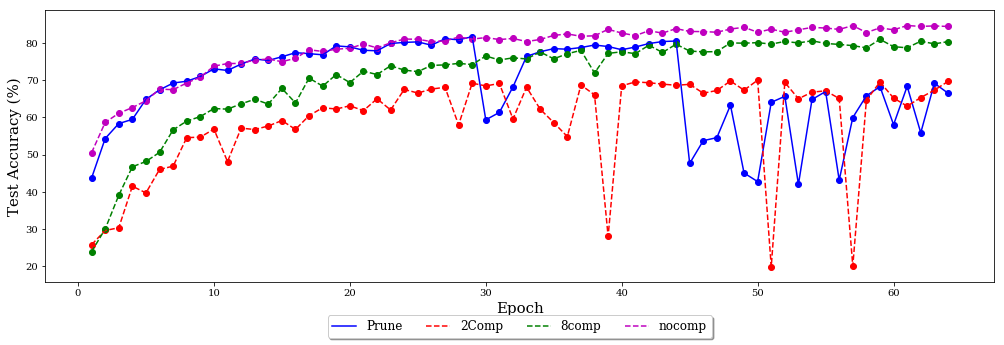}
  }
  \caption{Progession of training with prune method, for CIFAR10 dataset, when compression can get very high} 
  \label{fig:prune_highComp}
\end{figure}

As we are able to observe in figure \ref{fig:prune_cifar10r3}, we see a similar improvement at speeding up training with prune method, as was the case in figure \ref{fig:prune_cifar10progress} in subsection \ref{subsec:evalprune}, with the difference here being that the compression method is also complemented with a resolution compression factor of 3x3 using convolution method discussed in \ref{subsec:formulation}. Here again, the maximum value for compression by 4 times is reached within 4 epochs for our method, but takes 34 epochs for 'training from scratch' (32 comp). The results are similar for compression factor of 32. We were able to observe these over-performances consistently. We did note that while for a lot of cases the obtained accuracy was similar albeit with different learning speeds, as is a known case for deep learning training methods, there is uncertainty and variability: sometimes our method gave accuracy levels different than 'training from scratch'. And sometimes the learning could be slower towards reaching the desired accuracy and sometimes faster; but such variances are expected in learning tasks like deep learning for vision. As shown in figure \ref{fig:prune_highComp}, in case of compression by 64 times for CIFAR10 dataset, and resolution compression by 3x3 times, both method, our method as well as training from scratch gave output where accuracy varied noticeably across the epochs. As a future potential direction of research, we could find a way to automatically explore right and minimum compression possibilities.


\end{document}